%% file: wz_trans_mom_dist.tex
\documentstyle[12pt]{article}

\input wz_trans_mom_dist_defs

\begin{document}

\begin{flushright}
\mbox{
\begin{tabular}{l}
    FERMILAB-PUB-97/207-T
\end{tabular}}
\end{flushright}
\vskip 1.5cm
\begin{center}
\Large
{\bf $W$ and $Z$ transverse momentum distributions:
 resummation in $\qt$-space}

\vskip 0.7cm
\large
R.K. Ellis and Sini\v{s}a Veseli \\
\vskip 0.1cm
{\small Theory Group, Fermi National Accelerator Laboratory,
P.O. Box 500, Batavia, IL 60510} \\
\today
\end{center}
\thispagestyle{empty}
\vskip 0.7cm

\begin{abstract}
We describe an alternative approach to the prediction of $W$ and $Z$
transverse momentum distributions based on an extended version of the DDT
formula. The resummation of large logarithms, mandatory at small $\qt$, is
performed in $\qt$-space, rather than in the impact parameter $b$. The
leading, next-to-leading and next-to-next-to-leading towers of logarithms
are identical in the $b$-space and $\qt$-space approaches. We argue that
these terms are sufficient for $W$ and $Z$ production in the region in
which perturbation theory can be trusted. Direct resummation in $\qt$-space
provides a unified description of vector boson transverse momentum
distributions valid at both large and small $\qt$.
\end{abstract}

\newpage

\section{Introduction}

We re-examine the transverse momentum distributions of
vector bosons, in view of the large data samples expected at
the Tevatron. In $p \bar{p}$ collisions at $\sqrt{S}=1.8~\TeV$
we expect about $10^5$~$W$ bosons and $10^4$~$Z$ bosons,
observed through their leptonic decays,
per $100~\mbox{pb}^{-1}$ of accumulated data. These events
will be invaluable for QCD studies, as well as for precision measurements of
the $W$ mass. In order to exploit these data samples fully
the experimenters will require detailed information
about the expected rapidity and transverse
momentum distributions of the vector bosons and of their decay products.

In QCD a vector boson acquires transverse
momentum $\qt$ by recoiling against one or more emitted
partons~\cite{FM,ESW}. Order by order in perturbation theory we encounter
logarithms, $\ln Q^2/\qt^2$, where $Q$ is the mass
of the lepton anti-lepton pair resulting from the vector boson decay.
These logarithms must be resummed to give an
accurate prediction in the low $\qt$ region. The original
approach to the summation of logarithms at small $\qt$ was provided
by Dokshitzer, Dykanov and Troyan (DDT) ~\cite{DDT} who derived an expression
(reproduced here for the case of massive photon production),
\beq
\frac{d \sigma}{d Q^2 d \qt^2 dy} = \frac{\sigma_0}{Q^2}\sum_q e_q^2
\frac{d}{d \qt^2}
\Bigg\{
f_{q/A}(\xa,\qt)f_{\bar q/B}(\xb,\qt)
\exp [\cT_{\mbox{\tiny DDT}}(\qt,Q)] + (q\lra \bq)
\Bigg\}\ ,
\label{orig_ddt}
\eeq
where $\cT$ is a leading log
Sudakov form-factor,\footnote{Other
notation will be defined in the body of the paper.}
\beq
\cT_{\mbox{\tiny DDT}}(\qt,Q) = -\int_{\qt^2}^{Q^2}
{d {\bar \mu}^2\over {\bar \mu}^2}
\frac{\as(\bar \mu)}{\pi} \frac{4}{3}
\Big(\ln \frac{Q^2}{\bar \mu^2} -\frac{3}{2} \Big) \ .
\eeq

The present state of the art in the theoretical description of
vector boson production is based on the $b$-space formalism
where $b$ is the impact parameter which is Fourier conjugate to
the vector boson transverse momentum.
The $b$-space formalism, which allows the implementation of
transverse momentum conservation for the emitted gluons,\footnote{See,
for example, Ref.~\cite{EFS}.} has the remarkable consequence that
the cross section at $\qt=0$ is calculable for very large
$Q$~\cite{PP,CSS}.\footnote{Very massive vector bosons are produced
at $\qt=0$ in association with semi-hard gluons which have zero
net transverse momentum. Unfortunately, in $W$ and $Z$ production
$Q$ is too low and this asymptotic regime does not apply.}

Nevertheless, in practice the $b$-space formalism has certain
disadvantages. Since the cross section is given as a Fourier
integral in $b$ which extends from $0$ to $\infty$,
one cannot make theoretical predictions for {\em any} $\qt$
without having a prescription
for dealing with the non-perturbative region of large $b$.
This problem can be solved by introducing an additional non-perturbative
form factor (to be determined from experiment), but that also leads to
unphysical behaviour of the cross section at large $\qt$, where
one should recover the ordinary perturbation theory result. These
points will be further discussed later in the text.

Clearly, if one could perform the Fourier integral in $b$
analytically and thus obtain an expression for the cross section
in $\qt$-space, the above problems would be solved. A model
for the non-perturbative region would have to be introduced only
at the very lowest values of $\qt$, and
one would have a unified description of vector boson transverse momentum
distributions valid at both small and large $\qt$.

In this paper we present an approach to resummation in $\qt$-space,
which is based on an extended version of the DDT formula.
The $b$-space
formalism~\cite{CSS}-\cite{BY}
resums the contributions to the cross section from the
following towers of logarithms ($L= \ln Q^2/\qt^2$):
\bea
&\mbox{L}:  & \frac{1}{\qt^2} \as^j L^{2j-1} \ ,\nonumber \\
&\mbox{NL}: & \frac{1}{\qt^2} \as^j L^{2j-2} \ , \nonumber \\
&\mbox{NNL}: & \frac{1}{\qt^2} \as^j L^{2j-3}\ ,  \nonumber \\
&\mbox{NNNL}: & \frac{1}{\qt^2} \as^j L^{2j-4} \ .
\eea
Our extended DDT expression agrees with the $b$-space results
for all but the NNNL series. However,
for vector bosons with masses less than $\MZ$, we find that the NNNL series
is numerically unimportant for $\qt > 3 \GeV$. Furthermore,
a specific choice of coefficients in the $\qt$-space Sudakov form factor
allows us to absorb the first term in the NNNL tower of logarithms, and to
obtain exact agreement with resummation in $b$-space to $\cO(\as^2)$.

Based on these results, we argue
that the $\qt$-space approach preserves
almost all the reliable features of the
$b$-space formalism,\footnote{The subleading terms in
$b$-space have a profound effect at
$\qt=0$. However, for $W$ and $Z$ production this region is dominated by
non-perturbative effects.} and that it also has certain practical advantages:
\begin{itemize}
\item
We avoid numerical pathologies in the matching, caused
by combining results from $b$-space and $\qt$-space.
Although the matching is formally included in the
$b$-space method~\cite{AEGM,AK,BQY,ERV},
the cross section is not correctly calculated
for $\qt \geq Q/2$. The cross section in this region is the result
of a delicate cancellation between the resummed and finite pieces.
The slightly different treatment of the two terms is sufficient to upset
the cancellation.
In contrast, the matching works well
in $\qt$-space, leading to a unified description of the
$\qt$ and $y$ distributions valid for all $\qt$.
\item
We need to introduce a model only at the
very lowest values of $\qt$.
\item
We have the practical advantage that we
avoid both the numerical Fourier transform and
multiple evaluations of the structure functions at each value of $\qt$.
\end{itemize}
A complete explanation of these
points will be found later in the paper.

It is important to emphasize here we are not challenging the
{\it theoretical} importance of the $b$-space formalism, which leads to
interesting results, particularly about the production of very
massive bosons.  Nevertheless, it is our opinion that in practice the
extended DDT approach is sufficient for the theoretical
description of the $W$ and $Z$ production.

The rest of the paper is organized as follows:
in Section \ref{bspace}
we review the $b$-space resummation. In Section \ref{qtspace}
we derive an extended version of the DDT expression, which forms
the basis of our approach.
Section \ref{results} contains comparison of
the perturbative Sudakov form factors in the $\qt$-space and $b$-space
formalisms, and shows that in the region where the latter is reliable
it is essentially identical with the former.
We also present a prescription for dealing with the non-perturbative
region of low $\qt$, and compare our results with typical
$b$-space calculations. Our conclusions are given in
Section \ref{conc}, while Appendix \ref{saddle} contains
the saddle point evaluation of the $b$-space expression for
the cross section at $\qt=0$.

\section{Resummation formalism in $b$-space}
\label{bspace}

The general expression for the resummed differential
cross section for vector boson production
in hadronic collisions may be written in the form
\bea
{ d \sigma(AB \rar V(\rar {l {\bar l'}}) X )
\over dq^2_T \, dQ^2 \, dy \,  d\cos{\theta} \, d\phi}
 &=&
{1 \over 2^8 N \pi S} \,
{Q^2 \over (Q^2 - \MV^2)^2 + \MV^2 \GV^2} \nonumber \\
&\times &
\bigg[Y_r(\qt^2, Q^2, y, \theta)
 + Y_f(\qt^2, Q^2,y, \theta, \phi) \bigg]  \ .
\label{total}
\eea
In the above, $N=3$ is the number of colors,
$\sqrt{S}$ is the total hadron-hadron center-of-mass energy,
while $\theta$ and $\phi$ refer to the lepton polar and azimuthal angles
in the Collins-Soper (CS) frame~\cite{CS2}. The mass and width of the
vector boson are denoted by $\MV$ and $\GV$.
The functions $Y_r$ and $Y_f$ stand for the resummed and
finite parts of the cross section, respectively. As the details
of the finite part are not important for the subsequent
discussion, we review here only the resummed part,
and refer the reader to Ref.~\cite{ERV} for the
complete description of $\cO(\as)$ finite part.

The resummed part of the cross section is given as the Fourier integral
over the impact parameter $b$,\footnote{The prime on the sum in
Eq.~(\ref{resum}) indicates that gluons are excluded from the summation.}
\bea
Y_r(\qt^2, Q^2, y, \theta) &=&  \Theta{(Q^2-\qt^2)} {1\over 2\pi}
\int_{0}^{\infty} db\;b  \, J_{0}(\qt  b)
\sum_{a,b}{}^\prime
 F^{NP}_{a b } (Q,b,\xa,\xb) \nonumber \\
&\times &
W_{ab} (Q,b_*,\theta)
f'_{a/A}(\xa,\frac{b_0}{b_*})
f'_{b/B}(\xb,\frac{b_0}{b_*})
\ ,
\label{resum}
\eea
where the variables $\xa$ and $\xb$ are given in
terms of the lepton pair mass $Q$ and rapidity $y$ as
\beq
\xa = \frac{Q}{\sqrt{S}} \exp{(y)}\ ,\
\xb = \frac{Q}{\sqrt{S}} \exp{(-y)}\ .
\eeq

The modified parton structure functions in Eq.~(\ref{resum}), $f'$,
are related to the $\overline{MS}$ structure functions, $f$,
by a convolution
\beq
f'_{a/H} (\xa,\mu) = \sum_{c}
\int_{\xa}^{1} {d z \over z} \,
C_{ac}\left( {\xa \over z},\mu \right)
f_{c/H} \left( z, \mu \right)\ ,
\eeq
where ($a,b \neq g$)~\cite{DS}
\bea
C_{ab}(z,\mu) &=&
\delta_{ab} \Bigg\{\delta(1-z)
+\bas(\mu) C_F  \Big[ 1-z
+(\frac{\pi^2}{2}-4) \delta(1-z)\Big] \Bigg\} \ , \\
C_{ag}(z,\mu) &=& \bas(\mu) T_R
\Big[ 2 z( 1-z) \Big]\ .
\eea
Here we have introduced
\beq
\bas(\mu) = \frac{\as(\mu)}{2 \pi}\ ,
\eeq
while $C_F=4/3$ and $T_R=1/2$ are the usual colour factors.

The function $W$ can be expressed in terms of
the Sudakov form factor $\cS(b,Q)$ and is given by
\beq
W_{ab} (Q,b,\theta)  = H^{(0)}_{ab} (\theta) \exp{[\cS(b,Q)]} \ ,
\label{w}
\eeq
where $H^{(0)}$, which includes the angular dependence
of the lowest order cross section and coupling factors,
is defined in Appendix A of Ref.~\cite{ERV}.
The Sudakov form factor itself is given as~\cite{CSS}
\beq
\cS(b,Q) = -\int_{b_0^2/b^2}^{Q^2}
{d {\bar \mu}^2\over {\bar \mu}^2}
\left[ \ln\left({Q^2\over {\bar \mu}^2}\right)
A\big(\bas({\bar \mu})\big) +
B\big(\bas({\bar \mu})\big) \right] \ ,
\label{sudff}
\eeq
with $b_0=2 \exp(-\gamma_E)\approx 1.1229$.
The coefficients $A$ and $B$ are perturbation series in $\as$,
\beq
A(\bas) = \sum^\infty_{i=1} \bas^i \sudai\ , \
B(\bas) = \sum^\infty_{i=1} \bas^i \sudbi\ .
\label{ab}
\eeq
The first two coefficients in the expansion of $A$ and $B$
are known~\cite{DS,EMP}:
\bea
\sudau&=&2 C_F\ ,\nonumber \\
\sudad&=&2 C_F
\Big( N (\frac{67}{18}-\frac{\pi^2}{6})-\frac{10}{9} T_R n_f \Big)
\ ,\nonumber \\
\sudbu&=&-3 C_F\ ,\nonumber \\
\sudbd&=& C_F^2 \Big(\pi^2-\frac{3}{4}-12 \zeta(3)\Big)
      +C_F N \Big( \frac{11}{9} \pi^2-\frac{193}{12}+6 \zeta(3)\Big)
\nonumber \\
&+&C_F T_R n_f \Big(\frac{17}{3}-\frac{4}{9} \pi^2\Big)  \ .
\label{ABcoefs}
\eea

One of the main advantages of the $b$-space resummation formalism
is that the simple form for $\cS(b,Q)$ as given in Eq.~(\ref{sudff}),
remains valid to all orders in perturbation theory~\cite{CSS}.
In addition, as mentioned above, for very large
values of the vector boson mass the $b$-space formulae make
definite predictions for the
$\qt=0$ behaviour of the cross section~\cite{PP,CSS}.

Unfortunately, the practical implementation of the $b$-space formulae
presents some difficulties.
The $b$-space integral
in the Bessel transform in Eq.~(\ref{resum})
extends from $0$ to $\infty$,
which means that one has to find a
way to deal with the non-perturbative region where $b$ is large.
That problem is usually circumvented
by evaluating $W$ and the parton structure functions at
\beq
b_* = \frac{b}{\sqrt{1+(b/\blim)^2}}\ ,
\label{blimeqn}
\eeq
which never exceeds the cut-off value $\blim$, and also by introducing
an additional function $F^{NP}$, which represents the non-perturbative
(large $b$) part of the Sudakov form factor,
to be determined from experiment~\cite{CSS}.
This is usually done
by assuming a particular functional form for $F^{NP}$ which involves
several parameters that can be adjusted in order to give the best
possible description of experimental data.
The specific choice of the  functional form for $F^{NP}$
is a matter of debate~\cite{DWS,LY,ERV},
but we will not discuss it further here.
The point which we would like to emphasize here is that without introducing
$b_*$ and $F^{NP}$ one would not be able to make theoretical predictions
for {\em any} value of $\qt$, even in the large $\qt$ region where
perturbation theory is expected to work well.

\begin{figure}[p]
\vspace{12.0cm}
\includegraphics{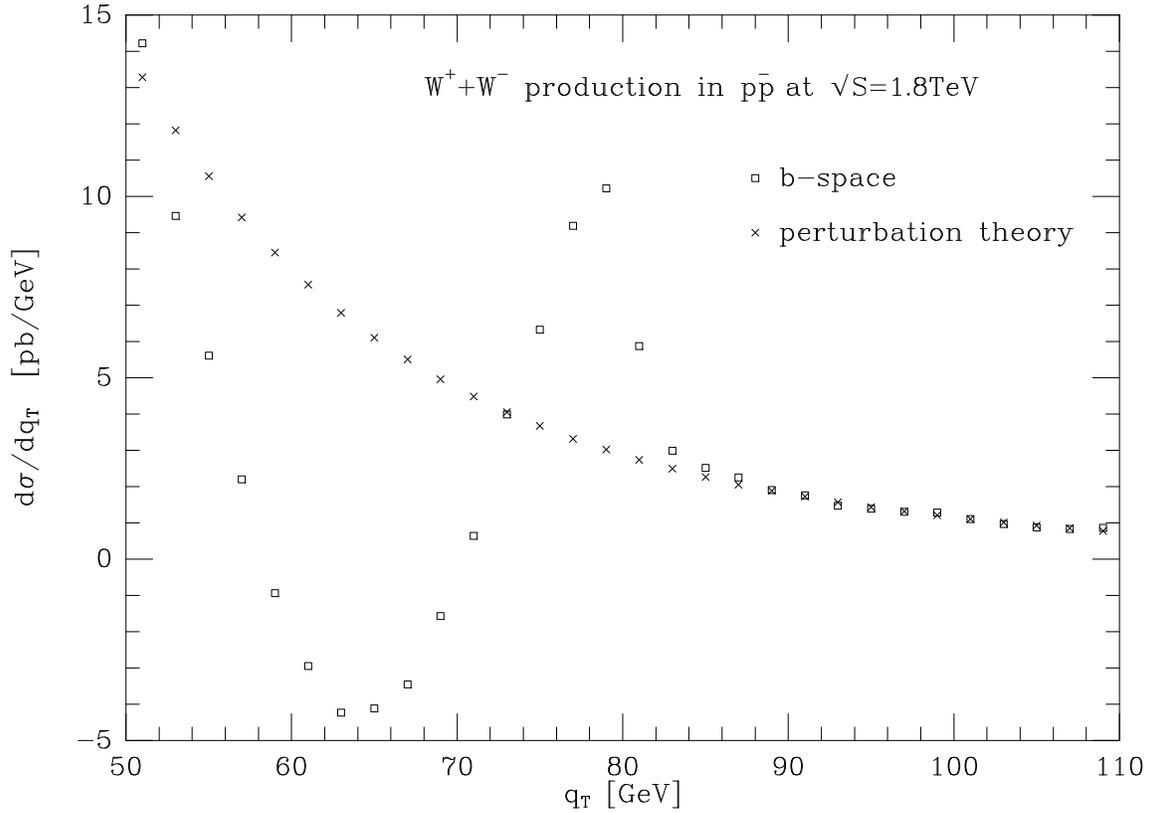}
\caption{Comparison of the $b$-space $d\sigma /d\protect \qt$ distribution
for $W^++W^-$ production at $\protect\sqrt{S}=1.8\protect\TeV$
with $\cO(\alpha_S)$ perturbative calculation. The resummation results
were obtained with pure gaussian ($g= 3.0\GeV^2,\protect\blim= 0.5\GeV^{-1}$)
form of $F^{NP}$. We assumed $BR(W\rar e\nu) = 0.111$.
}
\label{match_b}
\end{figure}

Another problem which occurs in the $b$-space resummation formalism
is the transition between the low and the high $\qt$ regions.
At large $\qt$ the resummed part is well represented by the first few terms in
its perturbative expansion. When the resummed part $Y_r$ is combined with
$Y_f$ one formally recovers the perturbation theory result.
However, the cancellation at large $\qt$ is quite delicate and
is compromised by the non-perturbative function which acts only on $Y_r$.
We illustrate the problem
in Figure \ref{match_b},\footnote{Note
that throughout the paper we use the MRSR1 structure
functions, with $\as(\MZ)=0.113$~\cite{MRS96}.}
which compares the $\cO(\as)$ perturbation
theory result for $d\sigma/d\qt$ in $W^++W^-$ production
at the Fermilab Tevatron,
to the theoretical prediction obtained
from the $b$-space resummation (Eqs.~(\ref{total}) and
(\ref{resum})).\footnote{Following Eqs.~(22,23) of Ref.~\cite{ERV},
instead of $b_0/b_*$ we actually used the exact first order result
for the scale at which parton distribution functions are evaluated.
This prescription preserves the total integral and reduces
to $b_0/b_*$ for large $b$. Furthermore, it
improves the large $\qt$ matching between $Y_r+Y_f$ and
the perturbation theory.}

Even though by carefully matching the low and high
$\qt$ regions one can reduce theoretical errors and
produce smoother transverse momentum distributions,
matching is still bound to fail eventually, and one is forced
to switch to the pure perturbative result at some $\qt$~\cite{AK}. This
procedure inevitably leads to discontinuous $\qt$ distributions,
which are clearly unphysical.

If one could find a $\qt$-space expression for $Y_r$,
both of the above problems would be solved: just as
for the conventional perturbation theory,
theoretical predictions could be made without
any smearing or additional functions, at least for values of $\qt$
not too close to zero. Also, since $Y_r$ and $Y_f$ would both
be calculated in $\qt$-space, the cancellation between the resummed
part and subtractions from the finite part would be explicit, and
matching of $Y_r+Y_f$ onto the perturbative result at large
$\qt$ would be manifest. With this motivation
we consider the derivation of $\qt$-space equivalent
of Eq.~(\ref{resum}) in the following section.

\section{Resummation in $\protect\qt$-space: extended DDT formula}
\label{qtspace}

For the sake of simplicity  we discuss only
the resummed part of the
non-singlet (NS) cross section for the process $AB \rar \gamma^* X $.
The extension to the general process $AB \rar V(\rar {l {\bar l'}}) X$
is straightforward.
In this case Eqs.~(\ref{total}),
(\ref{resum}) and (\ref{w}) can be rewritten in the form
\bea
{ d\sigma \over dq^2_T \, dQ^2}
 &=& {\sigma_0 \over Q^2} \sum_{q} e_q^2
\int_0^1d\xa d\xb\, \delta(\xa\xb-\frac{Q^2}{S})\nonumber \\
&\times&
 {1\over 2}
\int_{0}^{\infty} db\;b  \, J_{0}(\qt  b)\, \exp{[\cS(b,Q)]}
\,\tilde{f}'_{q/A}(\xa,\frac{b_0}{b})
\,\tilde{f}'_{\bq/B}(\xb,\frac{b_0}{b})
\ ,
\label{nscs}
\eea
where $\sigma_0 = 4 \pi \alpha^2 / (9S)$ and
$\tilde{f}'_{q/A}=f'_{q/A}-f'_{\bq/A},\,
\tilde{f}'_{\bq/B}=f'_{\bq/B}-f'_{q/B}$
are the higher order NS structure functions. Note that we have removed
the non-perturbative function $F^{NP}$ and variable $b_*$ from
Eq.~(\ref{resum}), so that the above expression
represents the pure perturbative result.

 From Eq.~(\ref{nscs}) one can easily obtain
the $N$-th moment of the cross section with respect to
$\tau = \xa\xb = Q^2/S$,
\bea
\Sigma(N) &=& \int d\tau\, \tau^N {Q^2 \over \sigma_0}
{ d \sigma \over dq^2_T \, dQ^2} \nonumber \\
&=&
\sum_{q} e_q^2
 {1\over 2}
\int_{0}^{\infty} db\;b  \, J_{0}(\qt  b)\, \exp{[\cS(b,Q)]}
\,\tilde{f}'_{q/A}(N,\frac{b_0}{b})
\,\tilde{f}'_{\bq/B}(N,\frac{b_0}{b})
\ .
\label{bsigma}
\eea
The $N$-th moment of the NS higher order structure function
satisfies the GLAP equation,\footnote{The anomalous dimension $\gamma^\prime$
differs in a calculable way from the $\overline{MS}$ anomalous dimension.}
\beq
\frac{d}{d \ln \mu^2 }f^\prime_{q/H}(N,\mu) =
\gamma^\prime_N f^\prime_{q/H}(N,\mu) \ ,
\eeq
with the solution
\beq
\tilde{f}'_{q/H}(N, \frac{b_0}{b}) =
\exp\left[
- \int_{(b_0/b)^2}^{Q^2}{d\bar{\mu}^2\over \bar{\mu}^2}
\gamma^\prime_N (\as(\bar{\mu}))\right]
\tilde{f}'_{q/A}(N,Q) \ .
\label{fns}
\eeq

Using Eqs.~(\ref{bsigma},\ref{fns}) we may write
\beq
\Sigma(N) = G(N,Q)
\, {1\over 2}
\int_{0}^{\infty} db\;b  \, J_{0}(\qt  b)\,
\,
\exp [\cU_N(b,Q)] \ ,
\label{sigma_j0}
\eeq
where $G(N,Q)$ denotes the parton flux,
\beq
G(N,Q) =  \sum_{q} e_q^2 \; \tilde{f}^\prime_{q/A}(N,Q)
 \tilde{f}^\prime_{\bar{q}/B}(N,Q) \ ,
\eeq
and the exponent $\cU$ is given as
\bea
\cU_N(b,Q) &=& -\int_{b_0^2/b^2}^{Q^2}
{d {\bar \mu}^2\over {\bar \mu}^2}
\left[
A\big(\bas({\bar \mu})\big) \ln {Q^2\over {\bar \mu}^2} +
B\big(\bas({\bar \mu})\big)
+2 \gamma_N^\prime  \big(\bas({\bar \mu})\big) \right] \nonumber \\
&\equiv& \sum_{n=1}^{\infty} \sum_{m=0}^{n+1} \bas^n(Q)
\ln^m\left({Q^2 b^2 \over b_0^2}\right) \dnm\ .
\label{exponent_in_dnm}
\eea
Here $\dud=-\frac{1}{2} \sudau$, etc.
Inserting  Eq.~(\ref{exponent_in_dnm}) in
Eq.~(\ref{sigma_j0}) we obtain
\beq
\Sigma(N) = G(N,Q)
\, {1\over 2}
\int_{0}^{\infty} db\;b  \, J_{0}(\qt  b)\,
\exp \left[
\sum_{n=1}^{\infty} \sum_{m=0}^{n+1} \bas^n(Q)
\ln^m\left({Q^2 b^2 \over b_0^2}\right) \dnm
\right]
\ .
\label{dnminb}
\eeq
This expression may be integrated by
parts using the relationship
\beq
\frac{d}{d x} \Big[x J_1(x)\Big] = x J_0 (x)\ .
\eeq
Because of the rapid damping of the
Sudakov factor as $b \rar \infty$ we may ignore the boundary terms
and obtain
\bea
\Sigma(N) &=&
-\frac{1}{2 \qt^2} G(N,Q)\,
\int_{0}^{\infty} dx \, J_{1}(x)\,
\frac{d}{dx}
\exp
\left[
\sum_{n=1}^{\infty} \sum_{m=0}^{n+1} \bas^n(Q)
\ln^m\left({Q^2 x^2 \over \qt^2 b_0^2}\right) \dnm
\right]
\ , \nonumber \\
& \equiv &
G(N,Q)\, \int_{0}^{\infty} dx \, J_{1}(x)\,
\frac{d}{d\qt^2}
\exp
\left[
\sum_{n=1}^{\infty} \sum_{m=0}^{n+1} \bas^n(Q)
\ln^m\left({Q^2 x^2 \over \qt^2 b_0^2}\right) \dnm
\right] \ .
 \label{DDTprototype}
\eea
Eq.~(\ref{DDTprototype})
already has the structure of the DDT formula. In fact, setting
$\ln x/b_0=0$ in the integrand we recover exactly the DDT formula,
i.e. the exponent has exactly the form of Eq.~(\ref{exponent_in_dnm})
with $b_0/b$ replaced by $\qt$. Because of that we write
$\Sigma(N)$ in the form
\bea
\Sigma(N)
 &=&
\frac{d}{d\qt^2} \Bigg\{
G(N,Q)\, \int_{0}^{\infty} dx \, J_{1}(x)\,
\exp \left[
\sum_{n=1}^{\infty} \sum_{m=0}^{n+1} \bas^n(Q)
\ln^m\left({Q^2 \over \qt^2 }\right) \dnm
\right] + R(\qt) \Bigg\} \nonumber \\
 &=&
\frac{d}{d\qt^2} \Bigg\{ G(N,Q)
\exp [\cU_N(\frac{1}{b_0 \qt},Q)]  + R(\qt)\Bigg\} \nonumber \\
 &\equiv&
\frac{d}{d\qt^2} \Bigg\{ G(N,\qt)
\exp [\cS(\frac{1}{b_0 \qt},Q)]  + R(\qt)\Bigg\} \ ,
\eea
where the remainder $R$ is defined as
\bea
R(\qt) &=&
G(N,Q)\, \int_{0}^{\infty} dx \, J_{1}(x)
\Bigg\{
\exp \left[ \sum_{n=1}^{\infty} \sum_{m=0}^{n+1} \bas^n(Q)
\ln^m\left({Q^2 x^2 \over \qt^2 b_0^2}\right) \dnm\right] \nonumber \\
&-&\exp \left[ \sum_{n=1}^{\infty} \sum_{m=0}^{n+1} \bas^n(Q)
\ln^m\left({Q^2  \over \qt^2 }\right) \dnm\right] \Bigg\} \ .
\eea
Using\footnote{See, for example, Ref.~\cite{EFS}.}
\beq
\int_0^{\infty} dx \, J_1(x) \,
\Big\{1,\ln \frac{x}{b_0},\ln^2 \frac{x}{b_0} ,
\ln^3 \frac{x}{b_0}, \ldots\Big\} =
\Big\{1,0,0,-\frac{1}{2} \zeta(3), \ldots \Big\}\ ,
\eeq
we can evaluate $R(\qt)$ as a power series in $\as$. We find that
the remainder contributes to the NNNL tower of terms,
three logarithms down from the leading terms ($L=\ln Q^2/\qt^2$),
\beq
R(\qt)=
 - G(N,Q) \Bigg\{ \zeta(3) \, \sum_{j=2}^\infty r_j ({}_1D_2 )^j
 \bas^j({Q}) L^{2j-3} +O(\bas^j L^{2 j-4}) \Bigg\}\ ,
\eeq
with
\beq
\Big\{r_2,r_3,r_4,r_5,r_6,r_7, \ldots\Big\} =
\Big\{8,\frac{40}{3},\frac{28}{3},4,\frac{11}{9},
\frac{13}{45},\ldots \Big\} \ .
\eeq

Starting from the $b$-space expression we have demonstrated
an extended DDT formula,
\bea
 { d\sigma \over dq^2_T \, dQ^2}
& =& {\sigma_0 \over Q^2} \sum_{q} e_q^2
\int_0^1d\xa d\xb\, \delta(\xa\xb-\frac{Q^2}{S})\nonumber \\
&\times&
{d \over d\qt^2} \Bigg\{ \,\tilde{f}'_{q/A}(\xa,\qt)
\,\tilde{f}'_{\bq/B}(\xb,\qt)
\, \exp{[\cT(\qt,Q)]}   + O(\bas^j L^{2 j-3}) \Bigg\}
\ ,
\label{qtnscs}
\eea
which holds if we drop NNNL terms.
In the above expression the $\qt$-space
Sudakov form factor is given by
\beq
\cT(\qt,Q) = -\int_{\qt^2}^{Q^2}
{d {\bar \mu}^2\over {\bar \mu}^2}
\left[
\tilde{A}\big(\bas({\bar \mu})\big)\ln{Q^2\over {\bar \mu}^2} +
\tilde{B}\big(\bas({\bar \mu})\big) \right]\ ,
\label{qtsudff}
\eeq
where the $\qt$-space coefficients $\tila$ and $\tilb$
are defined in a similar way as their $b$-space counterparts, i.e.
\beq
\tila(\as) = \sum^\infty_{i=1} \bas^i \tilai\ , \
\tilb(\as) = \sum^\infty_{i=1} \bas^i \tilbi\ .
\label{qtab}
\eeq
The first two coefficients in $\tila$ and $\tilb$ would be
exactly the same as corresponding $b$-space coefficients if
we drop NNNL terms. However, by
making the particular choice of
\bea
\tilau &=& \sudau \ , \nonumber \\
\tilad &=& \sudad \ , \nonumber \\
\tilbu &=& \sudbu \ , \nonumber \\
\tilbd &=& \sudbd + 2 (\sudau)^2 \zeta(3)\ ,
\label{qtabresult}
\eea
we absorb the first term in the NNNL tower of logarithms. In this way
Eq.~(\ref{qtabresult}) imposes exact agreement
between the $b$-space and $\qt$-space
formalisms at order $\as^2$.

As we pointed out at the beginning
of this section,
the extension of the NS cross section for $AB \rar \gamma^* X $
to the general case of $AB \rar V(\rar {l {\bar l'}}) X$ which
includes the decay presents no difficulties, so that
our $\qt$-space equivalent of Eq.~(\ref{resum})
is given in the extended DDT form as
\bea
\tilde{Y}_r(\qt^2, Q^2, y, \theta) &=&  \Theta{(Q^2-\qt^2)} {1\over \pi}
\nonumber \\
&\times&
\sum_{a,b}{}^\prime H^{(0)}_{ab} (\theta)
{d \over d\qt^2} \left[\, f'_{a/A}(\xa,\qt)
\,f'_{b/B}(\xb,\qt)
\, \exp{[\cT(\qt,Q)]}\,\right]
\ ,
\label{central_result}
\eea
with $\cT$ given in Eq.~(\ref{qtsudff}) in terms of coefficients
of Eqs.~(\ref{qtab},\ref{qtabresult}).
The above equation is the central result of this paper.
It is still ill-defined in the small $\qt$ region,
which reflects the fact that the problem is not entirely determined
by perturbation theory and requires non-perturbative input. We will discuss
our model for the non-perturbative region later in the following section.

\section{Results}
\label{results}

\subsection{Form factors}

Before presenting our results for $W$ and $Z$ production we compare
the form factors calculated using the $b$-space and $\qt$-space
formulae, for values of $Q$ which are presently of interest.  In
practice this means $Q \leq \MZ$.  The comparison of the form factors
will allow us to make an estimate of the practical numerical
importance of transverse momentum conservation, i.e. of the subleading
terms which are not present in the $\qt$-space formalism.  To simplify
the comparison we will consider the effects of the Sudakov form factor
alone. We will therefore ignore the influence of modified parton
distribution functions on the $\qt$ dependence.  For
the purpose of illustration we take $Q=\MZ$ and $\as(\MZ)=0.113$.

\begin{figure}[p]
\vspace{12.0cm}
\includegraphics{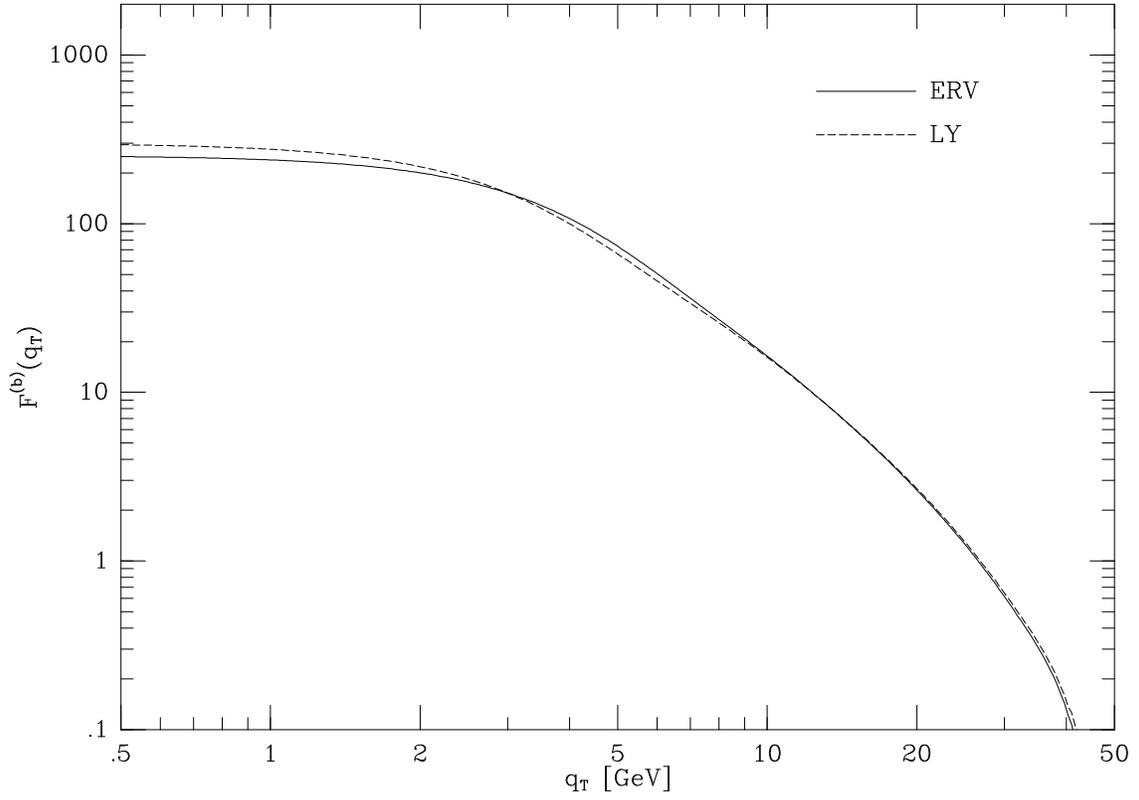}
\caption{{$F^{(b)}(\protect\qt)$} for the
two different choices of the non-perturbative function.}
\label{ff_bb}
\end{figure}

We define the $b$-space form factor as
\bea
F^{(b)}(\qt)&=& \frac{Q^2}{4 \pi } \int d^2b\;  \exp(i b\cdot \qt )\;
\exp[\cS(b_*,Q)]\; F^{NP}(Q,b,\xa,\xb) \nonumber \\
&=& \frac{Q^2}{2} \int_{0}^{\infty} db\, b \, J_0(b \qt)\; \exp[\cS(b_*,Q)]\;
 F^{NP}(Q,b,\xa,\xb)
\ .
\label{bff}
\eea
Note that $F^{NP}$ and $b_*$
have to be introduced in the above expression as a prescription for
dealing with the non-perturbative region of large $b$.
A specific choice of the non-perturbative function should make a difference
only in the region of low $\qt$. In order to show that, in Figure
\ref{ff_bb} we present form factors evaluated with
$F^{NP}$ taken from Ref.~\cite{LY} (LY), and with an  effective
gaussian as used in Ref.~\cite{ERV} (ERV, $g = 3.0\GeV^2$).
For LY form factor
we take $\xa=\xb=\MZ/\sqrt{S}$ for $\sqrt{S}=1.8~\TeV$.
As expected, at large $\qt$ the form factors resulting from
the two choices of $F^{NP}$ agree.
For small $\qt$ we find that results for
$F^{(b)}(\qt)$ tend to a different
finite intercept controlled by the
non-perturbative function.

\begin{figure}[p]
\vspace{12.0cm}
\includegraphics{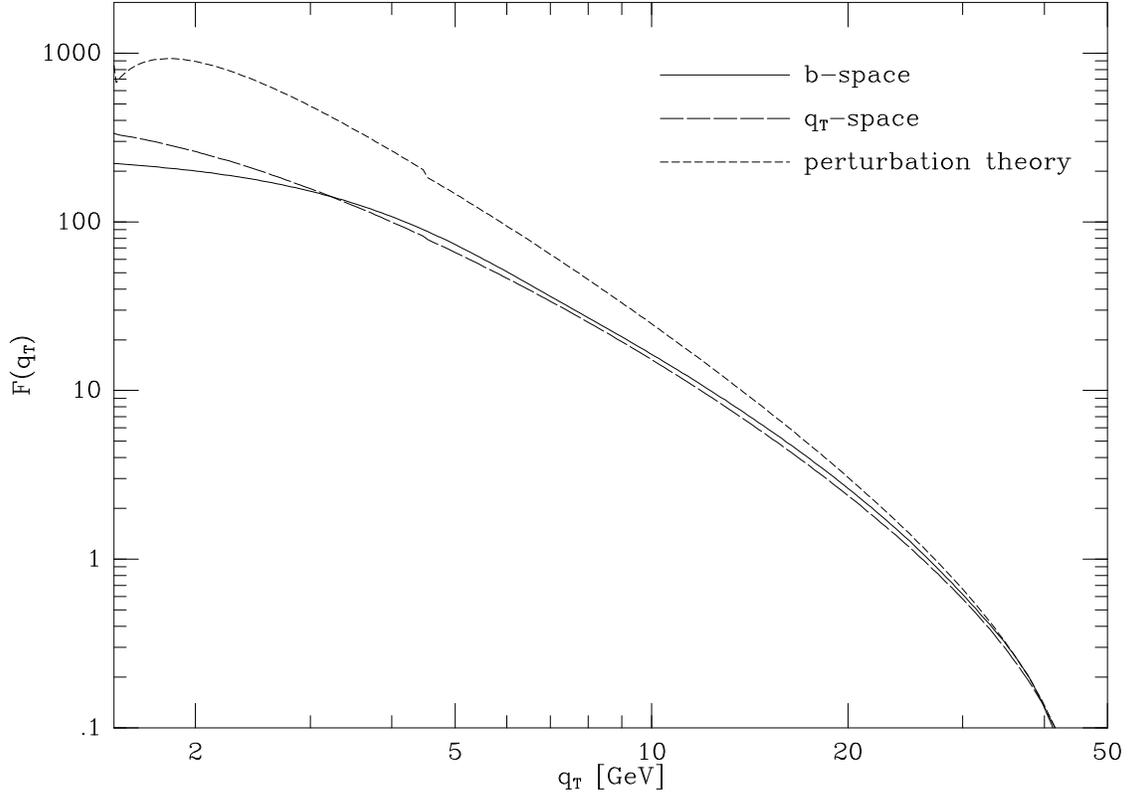}
\caption{Form factors $F^{(b)}$, $F^{(\protect\qt)}$ and
$F^{(p)}$. The $b$-space results were obtained with an effective
gaussian form of $F^{NP}$
($g = 3.0\protect\GeV^2,\protect\blim=0.5\GeV^{-1}$).}
\label{ff_bqt}
\end{figure}

The above $b$-space expression for the form factor
should be compared to its $\qt$-space counterpart,
\beq
F^{(\qt)}(\qt)= Q^2 \frac{d}{d \qt^2} \exp[\cT(\qt,Q)]\ ,
\label{qtff}
\eeq
and also to the $\cO(\as^2)$ perturbation theory result,
\beq
F^{(p)}(\qt)=
\frac{Q^2}{\qt^2}
\sum_{n=1}^{2}
\sum_{m=0}^{2 n-1} \bas^n
\ln^m \frac{Q^2}{\qt^2} \cnm \ ,
\label{pff}
\eeq
with $\cnm$ given in terms of the $\qt$-space coefficients as
\bea
\cuu&=& \tilau\ ,\nonumber \\
\cuz&=&  \tilbu\ ,\nonumber \\
\cdt&=&-\frac{1}{2} \left(\tilau\right)^2 \ , \nonumber \\
\cdd&=& \tilau (\beta_0  - \frac{3}{2}\tilbu)\ ,\nonumber \\
\cdu&=&  \tilad + \tilbu (\beta_0 - \tilbu)\ , \nonumber  \\
\cdz&=& \tilbd \ .
\eea
As one can see from Figure \ref{ff_bqt}, in the region
where one can trust perturbation theory
($\qt\geq 3\GeV$), our $\qt$-space
result of Eq.~(\ref{qtff}) agrees well with the $b$-space
form factor (obtained with ERV non-perturbative function).
Further, it is clear that resummation is needed in
the region where $F^{(b)}$ and $F^{(\qt)}$ differ significantly
from the perturbative result.

\begin{figure}[p]
\vspace{12.0cm}
\includegraphics{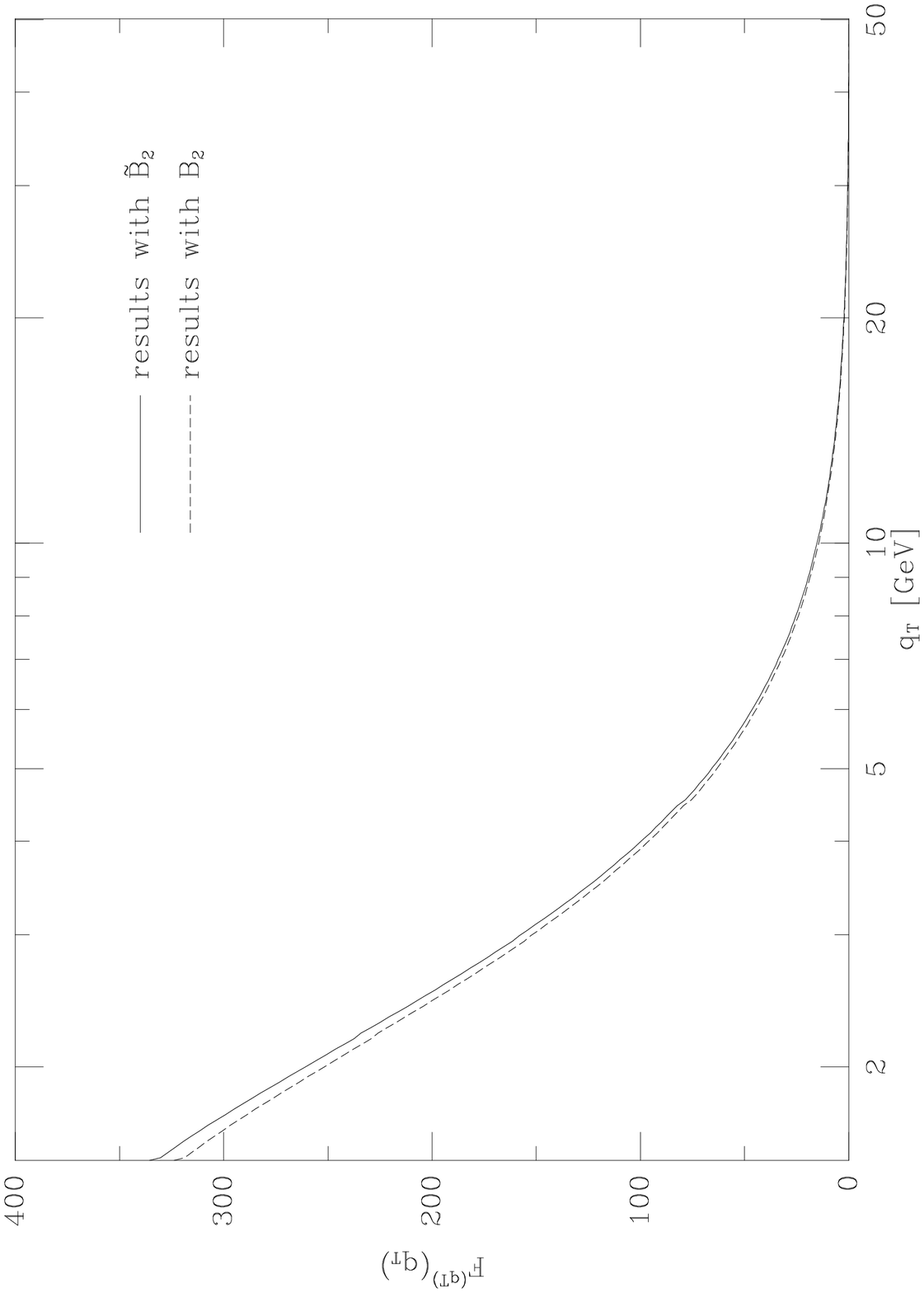}
\caption{The $\protect\qt$-space form factor $F^{(\protect\qt)}$
calculated with $\protect\sudbd$ and  $\protect\tilbd$.}
\label{b2effects}
\end{figure}

It is also interesting to investigate the size of the
NNNL effects.
In Figure \ref{b2effects} we show the $\qt$-space form factor
$F^{(\qt)}(\qt)$ calculated using coefficients
given in Eq. (\ref{qtabresult}), and also the
one calculated with $\tilbd$ replaced by $\sudbd$.
As one can see, the change is never more than a few
percent for $\qt > 3 \GeV$.

We therefore conclude that the $b$-space and $\qt$-space formula are
substantially identical, despite the neglect of NNNL terms in the latter.
The differences between them are smaller than the differences introduced
in the $b$-space formalism by the use of different non-perturbative
functions.  The above conclusion holds for the particular case of
the vector boson production with $Q \leq M_Z$.

\subsection{Extension to the non-perturbative region}

\begin{figure}[p]
\vspace{12.0cm}
\includegraphics{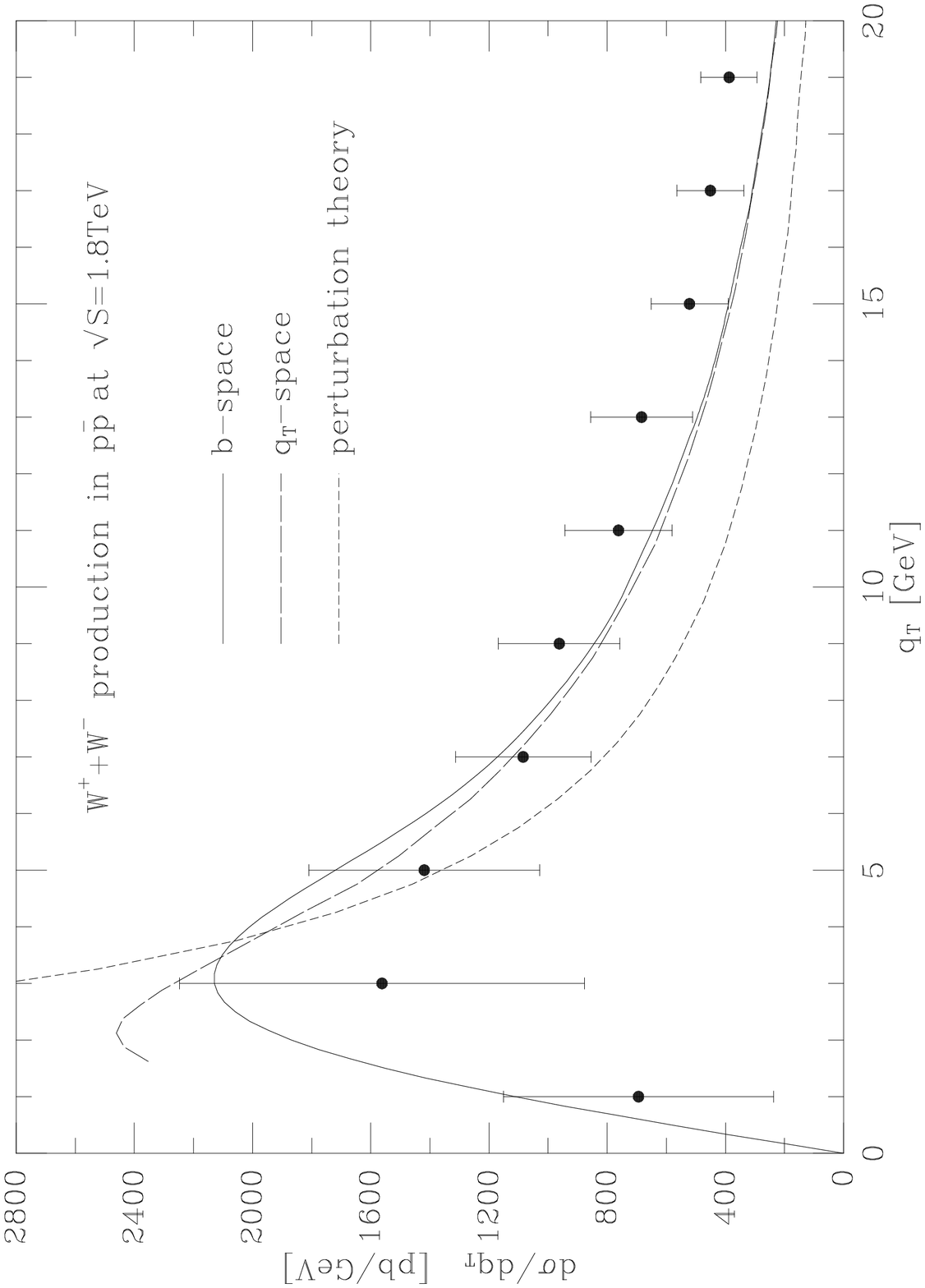}
\caption{Comparison of various theoretical predictions for
$W^++W^-$ $d\sigma /d\protect \qt$ with CDF
data \protect\cite{CDFW}. The $b$-space results were obtained with
an effective gaussian form of $F^{NP}$
($g = 3.0\protect\GeV^2,\protect\blim=0.5\GeV^{-1}$). We assumed
$BR(W\rar e\nu) = 0.111$.
}
\label{wpm_comp_b_qt}
\end{figure}

\begin{figure}[p]
\vspace{12.0cm}
\includegraphics{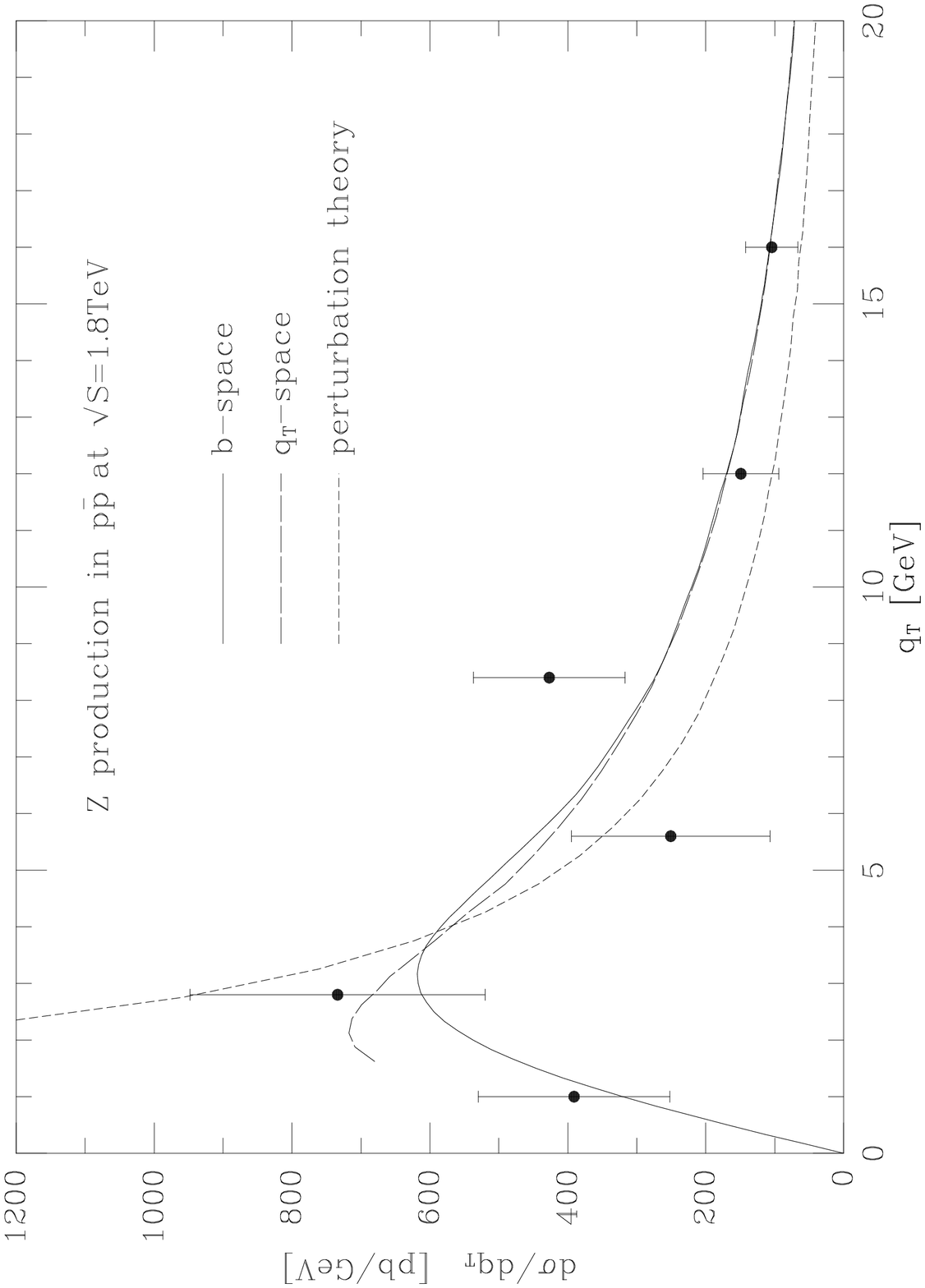}
\caption{Comparison of various theoretical predictions for
$Z$ $d\sigma /d\protect \qt$ with CDF
data \protect\cite{CDFZ}. The $b$-space
results were obtained with an effective gaussian form of $F^{NP}$
($g = 3.0\GeV^2,\blim=0.5\GeV^{-1}$). We assumed
$BR(Z\rar e^+e^-) = 0.033$.
}
\label{z0_comp_b_qt}
\end{figure}

\begin{figure}[p]
\vspace{12.0cm}
\includegraphics{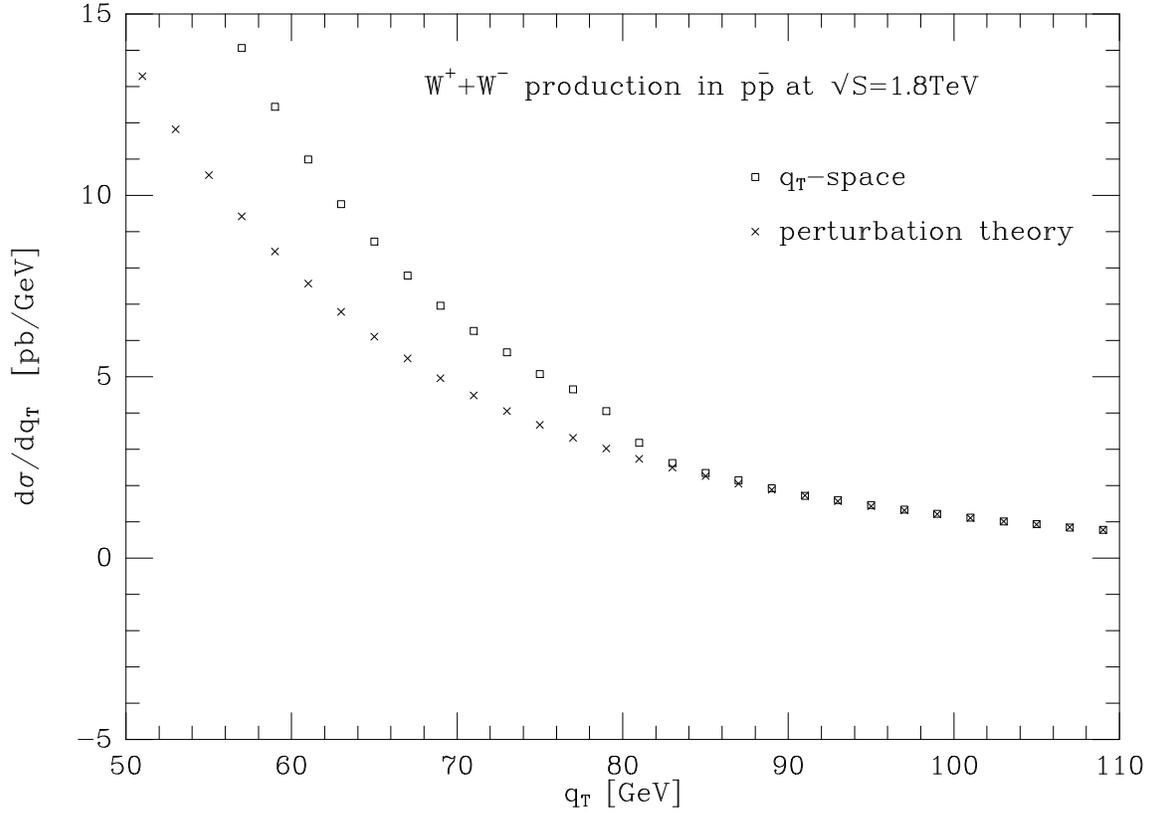}
\caption{Comparison of the $\protect\qt$-space
$d\sigma /d\protect \qt$ distribution
for $W^++W^-$ production at $\protect\sqrt{S}=1.8\protect\TeV$
with $\cO(\protect\as)$ perturbative calculation. We assumed
$BR(W\rar e\nu) = 0.111$.
}
\label{match_qt}
\end{figure}

As we have already pointed out, there are two main advantages
of the $\qt$-space approach  over the $b$-space
formalism: first, outside of the non-perturbative region
one can make theoretical predictions based on
perturbation theory alone with soft gluon resummation effects included.
In Figures \ref{wpm_comp_b_qt} and \ref{z0_comp_b_qt} we show
predictions of Eq.~(\ref{central_result}) for $W^++W^-$
and $Z$ production at Fermilab Tevatron. It is clear that
these predictions are quite close to typical $b$-space results.
Second, matching of the resummation formalism onto
pure perturbation theory for large $\qt$ is explicit, and hence
there is no need for somewhat unnatural switching from one
type of theoretical description to another. Since
our calculation contains the $\cO(\as)$
finite part and the $\cO(\as^2)$ Sudakov form factor, there
still may be some residual unmatched higher order effects present
in $d\sigma/d\qt$ in the large $\qt$ region, where the
cancellation of the resummed part and subtractions from
the finite part is quite delicate. However, these effects
are expected to be small, and should be
even less important after the inclusion of
the second order calculation of $Y_f$.
The $\qt$-space matching is illustrated
in Figure \ref{match_qt}
for $W^++W^-$ production at Tevatron, and
should be compared to the $b$-space result shown in Figure \ref{match_b}.
Note that less than $2\%$ of the total cross section lies above $\qt=50\GeV$,
so the overall importance of the portion of the cross section
shown in Fig.~\ref{match_qt} is quite small.

Up to now we have discussed only the $\qt$-space predictions
in the perturbative region, i.e. for $\qt \geq 2-3\GeV$.
Still, in order to compare theoretical predictions to
experiment one has to find a way of dealing with the non-perturbative
region ($\qt\rar 0$), where Eq.~(\ref{central_result}) is ill-defined.
The form of $\tilde{Y}_r$ suggests that we make
the following replacement in
Eq.~(\ref{central_result}):
\bea
&&f'_{a/A}(\xa,\qt)
\,f'_{b/B}(\xb,\qt)
\, \exp{[\cT(\qt,Q)]}\, \lar \nonumber \\
&&\hspace*{+2cm} f'_{a/A}(\xa,\qts)
\,f'_{b/B}(\xb,\qts)
\, \exp{[\cT(\qts,Q)]}\, \tilde{F}^{NP}(\qt)\ .
\eea
Here, $\qts$ is the effective transverse momentum
and $\tilde{F}^{NP}$ is the $\qt$-space non-perturbative part
of the form factor.
Since the above replacement should affect only the region of small
$\qt$, we
define $\qts$ as
\beq
\qts^2 = \qt^2 + \qtlim^2 \exp{\left[-\frac{\qt^2}{\qtlim^2}\right]}\ ,
\label{qtstar}
\eeq
which never goes below the limiting value $\qtlim$, and also
approaches $\qt$ as $\qt$ becomes much larger than $\qtlim$.
For $\tilde{F}^{NP}$ we require that
\bea
\tilde{F}^{NP}(\qt) &\rar &0 \hspace*{1.285cm}
({\rm for}\, \qt\rar 0)\ ,\nonumber \\
\tilde{F}^{NP}(\qt) &\rar &1 \hspace*{1.285cm}
({\rm for}\, \qt\rar Q)\ ,\nonumber\\
{d\over d\qt^2}\tilde{F}^{NP}(\qt) &\rar & const.\ \ \
({\rm for}\, \qt\rar 0)\ .
\eea
The first two properties ensure that
the integral of $\tilde{Y}_r$ over $\qt^2$ gives the result
\beq
\int_0^{Q^2} d \qt^2 \; \tilde{Y}_r
(\qt^2, Q^2, y, \theta) =  {1\over \pi}
\sum_{a,b}{}^\prime H^{(0)}_{ab} (\theta) f'_{a/A}(\xa,Q) f'_{b/B}(\xb,Q)
\ ,
\eeq
which is required to reproduce the exact $O(\as)$
total cross section after integration over $\qt$, as explained
in Ref.~\cite{ERV}.
The third condition is motivated by the analytic $b$-space results
for $d\sigma/d\qt^2$ in the limit where $\qt\rar 0$  ~\cite{PP}
(see Appendix \ref{saddle}).

\begin{figure}[p]
\vspace{12.0cm}
\includegraphics{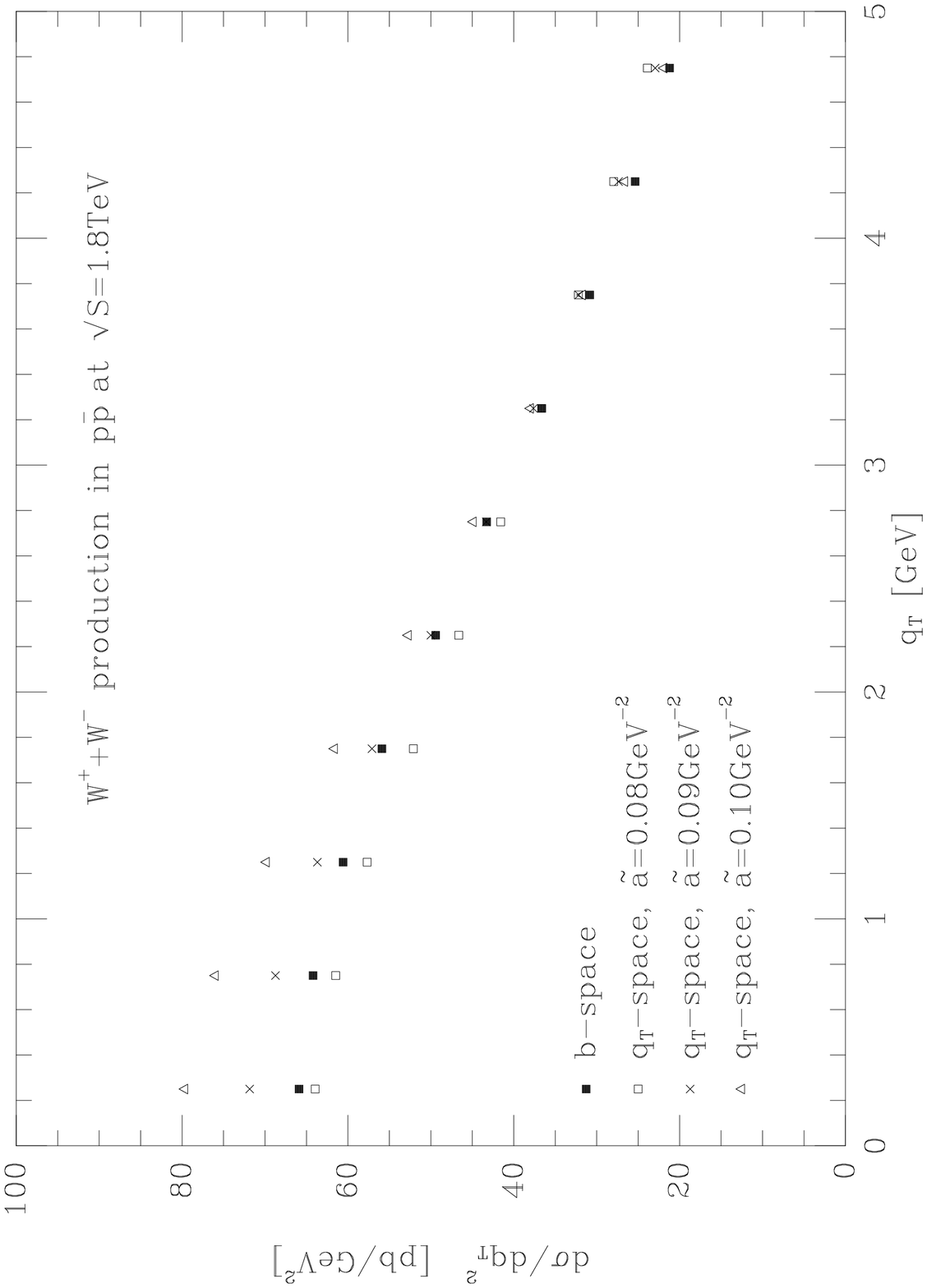}
\caption{Various theoretical predictions for
$d\sigma / d\protect\qt^2$ in
$W^++W^-$ production at $\protect\sqrt{S}=1.8\TeV$.
The $b$-space results were obtained with an effective gaussian form
of $F^{NP}$ ($g = 3.0\GeV^2,\blim=0.5\GeV^{-1}$). The
$\protect\qt$-space predictions correspond to $\protect\qtlim=4.0\GeV$.
}
\label{wpm_ds_dqt2}
\end{figure}

\begin{figure}[p]
\vspace{12.0cm}
\includegraphics{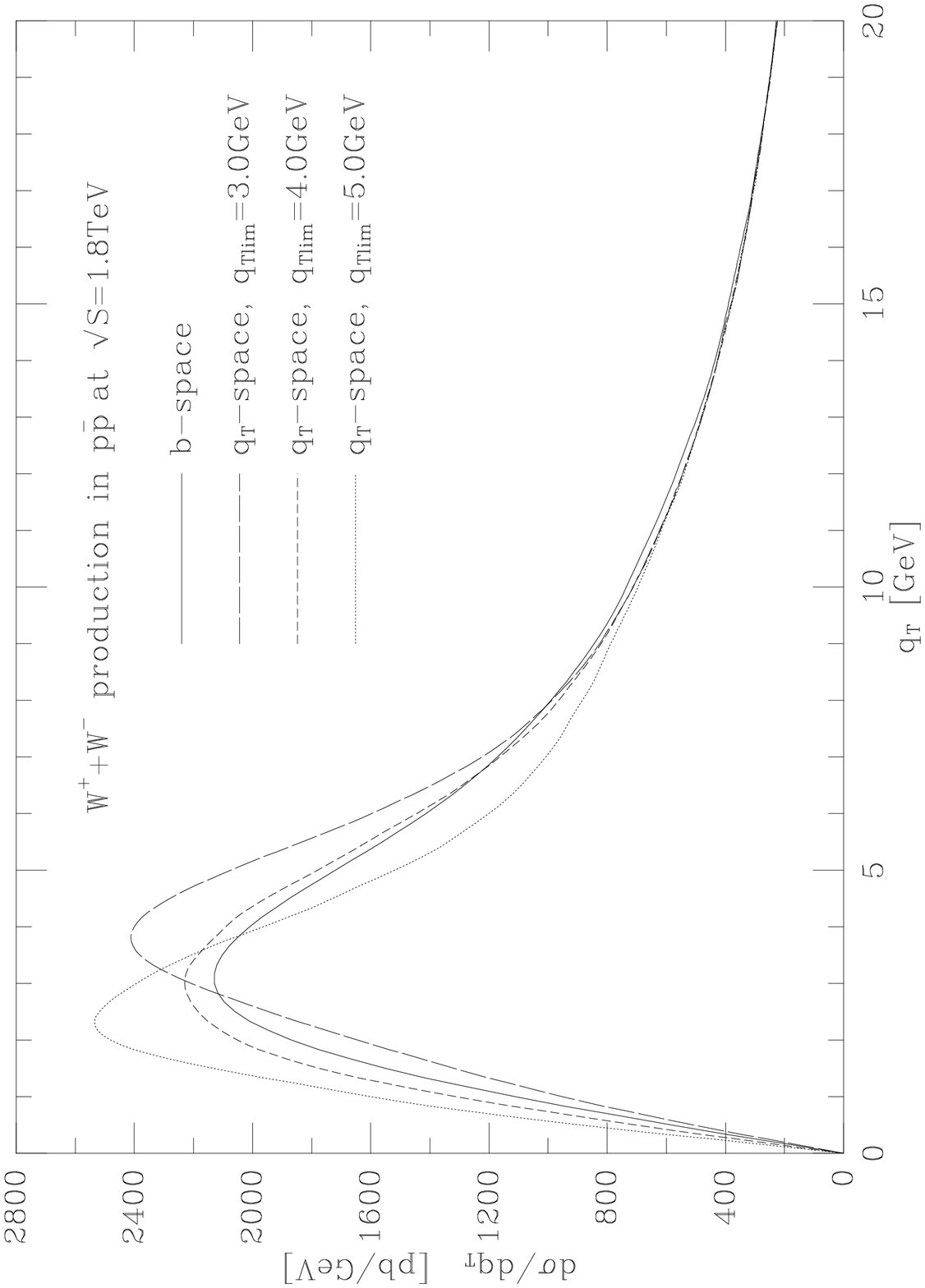}
\caption{Various theoretical predictions for
$d\sigma /d\protect \qt$ in $W^++W^-$ production
at $\protect\sqrt{S}=1.8\TeV$. The $b$-space
results were obtained with an effective gaussian form of $F^{NP}$
($g = 3.0\protect\GeV^2,\protect\blim=0.5\GeV^{-1}$).
The $\protect\qt$-space predictions correspond to
$\tilde{a}=0.10\protect\GeV^{-2}$. We assumed
$BR(W\rar e\nu) = 0.111$.
}
\label{wpm_comp_b_qt2}
\end{figure}

A simple choice for $\tilde{F}^{NP}$ which satisfies all of
the above requirements is
\beq
\tilde{F}^{NP}(\qt) = 1-\exp{[-\tilde{a}\, \qt^2]}\ .
\label{qtfnp}
\eeq
At $\qt=0$ this yields
\bea
{d\sigma \over d\qt^2} &\propto& \tilde{a}\,
\sum_{a,b}{}^\prime H^{(0)}_{ab} (\theta)
 \left[\, f'_{a/A}(\xa,\qtlim)
\,f'_{b/B}(\xb,\qtlim)
\, \exp{[\cT(\qtlim,Q)]}\,\right] \ , \nonumber\\
{d\over d\qt^2} \Big({d\sigma \over d\qt^2}\Big) &\propto& -\tilde{a}^2\,
\sum_{a,b}{}^\prime H^{(0)}_{ab} (\theta)
\left[\, f'_{a/A}(\xa,\qtlim)
\,f'_{b/B}(\xb,\qtlim)
\, \exp{[\cT(\qtlim,Q)]}\,\right] \ .\ \
\eea
Therefore, $\tilde{a}$
and $\qtlim$  control the intercept and the first derivative
of  $d\sigma/d\qt^2$ at $\qt=0$.
The effects of changing these non-perturbative parameters
are illustrated in Figures \ref{wpm_ds_dqt2} and \ref{wpm_comp_b_qt2},
for $W^++W^-$ production at Fermilab Tevatron.
In Figure \ref{wpm_ds_dqt2} we compare typical
$b$-space results for the  $d\sigma /d\qt^2$ distribution,
to the $\qt$-space predictions with several
different values of $\tilde{a}$ ($\qtlim$ was fixed to $4.0\GeV$).
In Figure \ref{wpm_comp_b_qt2} we plot
our $d\sigma/d\qt$ results obtained with
$\tilde{a}$ fixed to $0.10\GeV^{-2}$, and for several
different choices of $\qtlim$. These results show how varying $\qtlim$
modifies the width and shifts the peak of the $d\sigma/d\qt$ distribution.

 From Figures \ref{wpm_ds_dqt2} and \ref{wpm_comp_b_qt2}
it is also clear that $\tilde{a}$ and $\qtlim$ affect only the low
$\qt$ region, while for $\qt\geq 10\GeV$ we again obtain the extended
DDT result of Eq. (\ref{central_result}). Because of that,
determination of these parameters from the experimental data
should not be too difficult.

\subsection{Overall smearing}

Introduction of $\tilde{a}$ and $\qtlim$
allowed us to extend the validity of Eq.~(\ref{central_result})
beyond the perturbative region in $\qt$.
However, this may not be enough for a good description
of experimental data, and one may need additional degrees of
freedom for modelling the low $\qt$ region.\footnote{We remind
the reader that some choices of
the non-perturbative function in the $b$-space formalism involve
4-6 different parameters.} This can be achieved by choosing more complicated
functional forms for $\qts$ and $\tilde{F}^{NP}$ than the
ones we suggested in Eqs. (\ref{qtstar}, \ref{qtfnp}), or by
imposing an overall smearing on the theoretical
transverse momentum distributions. Here we briefly discuss the
later possibility.

\begin{figure}[p]
\vspace{12.0cm}
\includegraphics{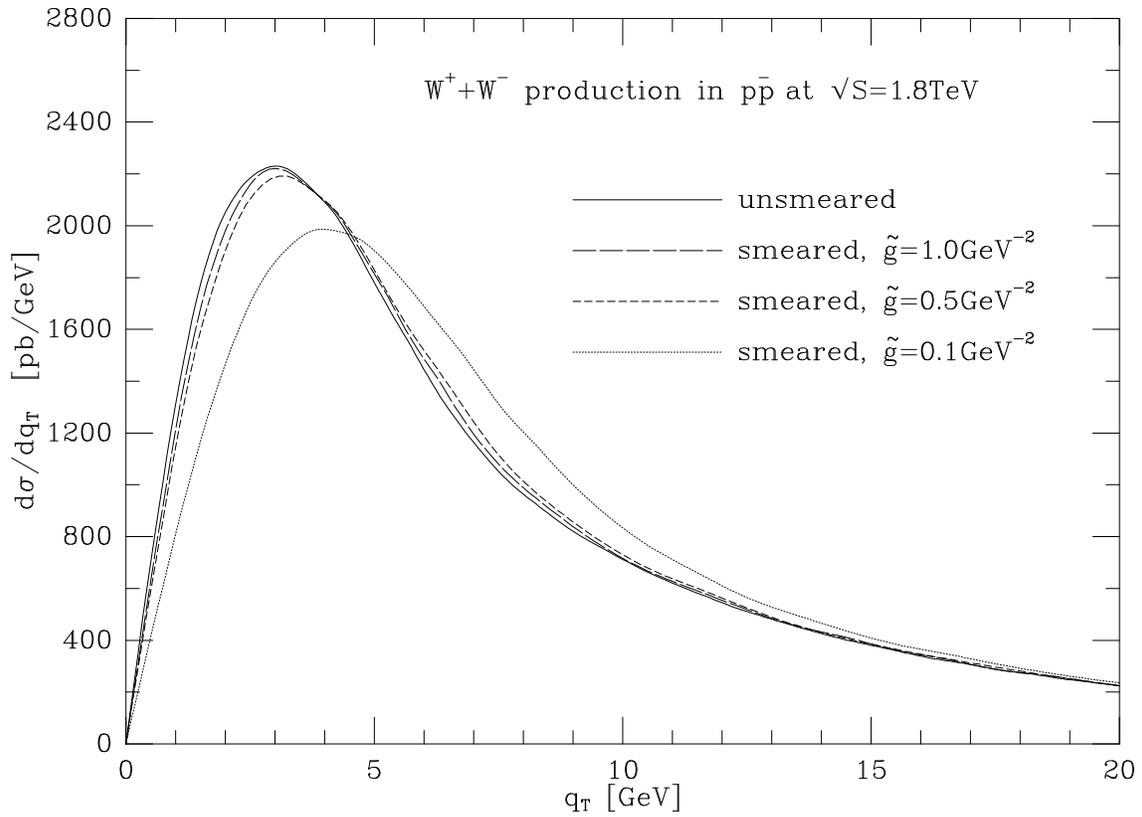}
\caption{Effects of smearing in $\protect\qt$-space for $W^++W^-$
production at $\protect\sqrt{S}=1.8\TeV$. We used
$\tilde{a}=0.10\protect\GeV^{-2}$, $\protect\qtlim=4.0\GeV$, and
$BR(W\rar e\nu) = 0.111$.
}
\label{wpm_comp_b_qt3}
\end{figure}

Suppressing irrelevant variables,
the smeared cross section is given in terms of
\beq
\tilde{Y}_i(\qt^2)
= \int \, d^2 \kt f(|\bkt-\bqt|) \,
\tilde{Y}_i(\kt^2)\ ,
\eeq
where $\tilde{Y}_i$ stands for either resummed or finite
part in $\qt$-space, and $f$ is the smearing function. For the sake of
simplicity we take a gaussian,
\beq
f(\kt) = \frac{\tilde{g}}{\pi}\exp( -\tilde{g}\, \kt^2)\ ,
\eeq
with $\tilde{g}$ being an additional non-perturbative
parameter. The above choice is convenient since
the azimuthal integration can be done analytically. This
leads to the final expression for the smeared $\tilde{Y}_i$,
\beq
\tilde{Y}_i( \qt^2)= \tilde{g} \int \, d \kt^2 \
\exp\Big[-\tilde{g}(\qt^2 +\kt^2) \Big]\, I_0 (2 \tilde{g} \qt \kt)\,
\tilde{Y}_i(\kt^2)\ .
\eeq
Note that both the resummed and the finite part of the cross section
can be smeared together, and therefore the smearing procedure should not
affect matching onto the pure perturbative result at large $\qt$.
The effects of an overall smearing for low $\qt$ are illustrated in
Figure \ref{wpm_comp_b_qt3} for $W^++W^-$ production
at Tevatron.

\section{Conclusions}
\label{conc}

In this paper we have outlined an approach to the calculation of
the transverse momentum distributions of $W$ and $Z$ bosons using an
extension of the DDT formula which works directly in $\qt$-space.
Our formalism agrees with $b$-space for all calculated logarithms
except the NNNL series.
This is a pragmatic approach which uses the available theoretical
in an efficient way. For $\qt$ above about $3\GeV$
the cross section is essentially determined by perturbative QCD.
In the region $\qt \leq 3\GeV$ the cross section is determined by a model,
the form of which is motivated by the analytic results from the $b$-space
approach. Just as in the $b$-space approach,
the details of the model are to be fixed by comparison
with experiment. The numerical program incorporating our results
describes all kinematic regions.

An obvious shortcoming of this paper is the failure to include the
results of the order $\as^2$ calculations~\cite{EMP,AR,GPW}
(generalized to include the decay
of the vector boson~\cite{M,GGK})
in the finite part of the cross section. In the $\qt$-space formalism
these should be relatively straightforward to include. After inclusion
of these effects we will have a full description of vector boson production
valid in all kinematic regions, with a minimum of model dependence.

\begin{center}
ACKNOWLEDGMENTS
\end{center}
This work was supported in part by the U.S. Department of Energy
under Contract No. DE-AC02-76CH03000.

\appendix

\section{Analytic behaviour at $\protect\qt=0$}
\label{saddle}

The result of Parisi and Petronzio~\cite{PP}
for the intercept at $\qt=0$
can be obtained by saddle
point evaluation of Eq.~(\ref{bff}),
\beq
F^{(b)}(0)  = \frac{Q^2}{4} \int d b^2 \exp [\cS(b,Q)] \ .\\
\eeq
In writing the above equation we have assumed that the saddle point value
of $b$ is small so that $b_*=b$ and $F^{NP}(b)=1$. Introducing the
variable $x=\ln b^2$  we have that
\beq
F^{(b)}(0)  = \frac{Q^2}{4} \int d x \exp [-h(x)]\ ,
\eeq
where
\beq
h(x)=-[x +\cS(\exp(x/2),Q)]\ .
\eeq
The saddle point result for $F^{(b)}(0)$ is then given by
\beq
F^{(b)}(0)
 = \frac{Q^2}{4} \sqrt{\frac{2 \pi}{h^{\prime \prime}(\xsp)}}
\exp [-h(\xsp)] \ ,
\eeq
where $\xsp=\ln \bsp^2$ is defined by the condition
\beq
h^\prime(\xsp)=0 \ .
\eeq
On the assumption that the structure functions are slowly varying functions
of the scale, the resummed part at $\qt=0$ becomes
\beq
Y_r(0, Q^2, y, \theta) =  {\bsp^2 \over 4\pi } \sqrt{\frac{2 \pi}
{-\cS^{\prime \prime}(\bsp,Q)}}
\sum_{a,b}{}^\prime
W_{ab} (Q,\bsp,\theta)
f'_{a/A}(\xa,\frac{b_0}{\bsp})
f'_{b/B}(\xb,\frac{b_0}{\bsp})
\ ,
\label{yrsp}
\eeq
where
\beq
\cS^{\prime \prime}(b,Q) = \frac{d^2 \cS(b,Q)}{d (\ln b^2)^2 }  \ .
\eeq

By retaining only the leading term ($A_1$) in the Sudakov form factor
we can obtain an
approximate analytic solution.
We assume that the running coupling satisfies the equation
($\beta_0 = (33-2 n_f)/6$)
\beq
\alpha_S(\mu)=\frac{2 \pi}{\beta_0} \frac{1}{\ln \mu^2/\Lambda^2}
\ ,
\eeq
and set $C=2 C_F/\beta_0$.
The saddle point of the integral is then given by
\beq
\frac{1}{\bsp}= \frac{\Lambda}{b_0}
\Big(\frac{Q}{\Lambda}\Big)^{\frac{C}{C+1}} \ .
\label{bsp}
\eeq
Using Eqs. (\ref{yrsp},\ref{bsp}) we obtain the final
result for the resummed part of the cross section ($L=\ln Q^2/\Lambda^2$),
\beq
Y_r(0, Q^2, y, \theta) \approx
\frac{b_0^2}{4 \pi \Lambda^2}
\frac{\sqrt{2 \pi C L}}{(C+1)}
\Big(\frac{\Lambda^2}{Q^2}\Big)^\eta
\sum_{a,b}{}^\prime
 H_{ab}^{(0)}(\theta)
f'_{a/A}(\xa,\frac{b_0}{\bsp})
f'_{b/B}(\xb,\frac{b_0}{\bsp}) \ ,
\eeq
with
\beq
\eta= C \ln \frac{C+1}{C} \ .
\eeq
The $Y_r$ has a finite intercept at $\qt=0$ which shrinks with $Q$.
For $n_f=3\, (4)$ we have that $\eta=0.586\, (0.602)$.

\clearpage
\newpage

\end{document}

%% file: wz_trans_mom_dist_defs.tex
\def\beq{\begin{equation}}
\def\eeq{\end{equation}}

\def\bea{\begin{eqnarray}}
\def\eea{\end{eqnarray}}
\def\rar{\rightarrow}
\def\lar{\longrightarrow}
\def\lra{\leftrightarrow}

\def\GeV{{\;\rm GeV}}

\def\TeV{{\;\rm TeV}}
\def\blim{b_{\;\rm lim}}
\def\as{\alpha_S}
\def\bas{\bar{\alpha}_S}

\def\bsp{b_{\mbox{{\tiny SP}}}}
\def\xsp{x_{\mbox{{\tiny SP}}}}
\def\GV{\Gamma_V}
\def\MV{M_V}
\def\MZ{M_Z}

\def\cO{{\cal O}}

\def\cS{{\cal S}}
\def\cT{{\cal T}}
\def\cU{{\cal U}}

\def\bq{\bar{q}}

\def\xa{x_{\mbox{\tiny A}}}
\def\xb{x_{\mbox{\tiny B}}}

\def\qt{q_{\mbox{\tiny T}}}
\def\qts{q_{\mbox{\tiny T}*}}
\def\qtlim{q_{\mbox{\tiny Tlim}}}
\def\kt{k_{\mbox{\tiny T}}}
\def\bqt{{\bf q}_{\mbox{\tiny T}}}
\def\bkt{{\bf k}_{\mbox{\tiny T}}}

\def\sudai{A^{(i)}}
\def\sudau{A^{(1)}}
\def\sudad{A^{(2)}}

\def\sudbi{B^{(i)}}
\def\sudbu{B^{(1)}}
\def\sudbd{B^{(2)}}

\def\tila{\tilde{A}}
\def\tilai{\tilde{A}^{(i)}}
\def\tilau{\tilde{A}^{(1)}}
\def\tilad{\tilde{A}^{(2)}}

\def\tilb{\tilde{B}}
\def\tilbi{\tilde{B}^{(i)}}
\def\tilbu{\tilde{B}^{(1)}}
\def\tilbd{\tilde{B}^{(2)}}

\def\dnm{{\;}_nD_m}
\def\dud{{\;}_1D_2}

\def\cnm{{\;}_nC_m}

\def\cuu{{\;}_1C_1}
\def\cuz{{\;}_1C_0}
\def\cdt{{\;}_2C_3}
\def\cdd{{\;}_2C_2}
\def\cdu{{\;}_2C_1}
\def\cdz{{\;}_2C_0}

\def\pl#1#2#3{
        {\it Phys.\ Lett.\ }{\bf #1}, #2 (19#3)}

\def\prl#1#2#3{
        {\it Phys.\ Rev.\ Lett.\ }{\bf #1}, #2 (19#3)}

\def\prep#1#2#3{
        {\it Phys.\ Rep.\ }{\bf #1}, #2 (19#3)}
\def\pr#1#2#3{
        {\it Phys.\ Rev.\ }{\bf #1}, #2 (19#3)}
\def\np#1#2#3{
        {\it Nucl.\ Phys.\ }{\bf #1}, #2 (19#3)}